\documentclass[%
 reprint,
 showpacs,
 showkeys,
 amsmath,amssymb,
 aps,
 prl,
]{revtex4-1}

\usepackage[english]{babel}
\usepackage{graphicx}
\usepackage{dcolumn}
\usepackage{bm}

\usepackage{siunitx}
\usepackage{float}
\usepackage{xcolor}

\begin{document}

\preprint{APS/123-QED}

\title{Pump-probe Spectroscopy Study of Ultrafast Temperature Dynamics in Nanoporous Gold}

\author{Michele Ortolani$^1$}
\author{Andrea Mancini$^1$}%
\author{Arne Budweg$^2$}%
\author{Denis Garoli$^3$}%
\author{Daniele Brida$^{2,4}$}%
\author{Francesco de Angelis$^3$}%

\affiliation{%
 $^1$ Department of Physics Sapienza University of Rome-00185 Rome-Italy. \\
 $^2$ Department of Physics and Center for Applied Photonics University of Konstanz-78457 Konstanz-Germany.  \\
 $^3$ Plasmon Nanotechnology Department Istituto Italiano di Tecnologia (IIT)-16163 Genoa-Italy.\\
 $^4$ Physics and Materials Science Research Unit University of Luxembourg-L-1511 Luxembourg-Luxembourg.\\
}%

\date{\today}

\begin{abstract}
We explore the influence of the nanoporous structure on the thermal 
relaxation of electrons and holes excited by ultrashort laser pulses ($\sim \SI{7}{fs}$) in thin gold 
films. Plasmon decay into hot electron-hole pairs results in the generation of a Fermi-Dirac distribution thermalized at a temperature $T_{\mathrm{e}}$ higher than the lattice temperature $T_{\mathrm{l}}$. The relaxation times of the energy exchange between electrons and lattice, here measured by pump-probe spectroscopy, is slowed down by the nanoporous structure, resulting in much higher peak $T_{\mathrm{e}}$ than for bulk gold films. 
The electron-phonon coupling constant and the Debye temperature are found to scale with the metal filling factor $f$ and a two-temperature model reproduces the data. The results open the way for electron temperature control in metals by engineering of the nanoporous geometry.
\end{abstract}

\maketitle



The optical excitation of electrons and holes at high energy levels in metal 
nanostructures has been the subject of considerable attention in the 
last decade \cite{atwater2010plasmonics,manjavacas2014plasmon,brongersma2015plasmon,wu2015efficient,benz2016single,cortes2017plasmonic}, with the aim of enabling chemical reactions and charge 
transfer from the metal to adjacent materials at ambient temperature for 
energy harvesting and storage \cite{atwater2010plasmonics,wu2015efficient}, most notably H$_2$ production by water splitting \cite{lee2012plasmonic,mukherjee2012hot,zhang2017surface,zhang2018plasmonic}. 
In particular, gold nanostructures have 
been investigated because of the relative ease of obtaining plasmonic 
field enhancement at their surfaces \cite{naik2013alternative}. The absorption of optical energy by free carriers in a metal implies collective oscillation of electron currents (plasmons) \cite{ruello2015ultrafast,Govorov,deacon2017interrogating}. 
Such coherent plasmons rapidly decay into non-thermalized electron-hole (e-h) pairs occupying high kinetic energy states. The e-h pairs decay via electron-electron scattering on the femtosecond time scale into hot carriers, which can be represented by a Fermi-Dirac distribution at an electron temperature $T_{\mathrm{e}}$, much higher than the lattice temperature $T_{\mathrm{l}}$. Subsequently, electron-phonon interaction leads to equilibrium defined as $T_{\mathrm{e}} \approx T_{\mathrm{l}}$ on the picosecond timescale \cite{baida2011ultrafast,della2012real}. 

Very recently, \emph{ab-initio} calculations of all electron and phonon states of gold have been employed to confirm the above interpretation of ultrafast pump-probe spectroscopy in the case of spherical nanoparticles of \SI{60}{nm} diameter in aqueous solution \cite{AtwaterPRL2017}. For such a simple geometry, electron and phonon distributions may be taken as constant in space, and the introduction of statistical thermal baths for electrons at $T_{\mathrm{e}}$ and phonons at $T_{\mathrm{l}}$ may not be conceptually necessary any more. The present work, however, explores the opposite limit of an extended nanoscale filament network, also called nanoporous gold (NPG). In NPG, geometrical parameters such as gold filling factor and filament diameter play a key role in determining the electron-phonon thermalization time due to spatially inhomogeneous excitation intensity at the nanoscale, therefore the previous simplified approach of two coupled statistical thermal baths (so called two-temperature (TT) model \cite{anisimov1974electron,carpene2006ultrafast,della2012real}) will be followed in this work so as to effectively include the geometrical parameters of the nanoporous gold structure in the model.

Hot electron plasmonics experiments have been mostly conducted on nanoparticles dispersed in solutions \cite{hodak1998spectroscopy,hrelescu2010dna,lee2012plasmonic,mukherjee2012hot,aruda2013identification,manjavacas2014plasmon,zhang2017surface,zhang2018plasmonic}, and the ultrafast temperature dynamics are poorly understood 
due to an extremely varied experimental landscape \cite{wu2015efficient,keller2018ultrafast,yu2018hot}.

\begin{figure}[ht]
\centering
\includegraphics[width=0.45\textwidth]{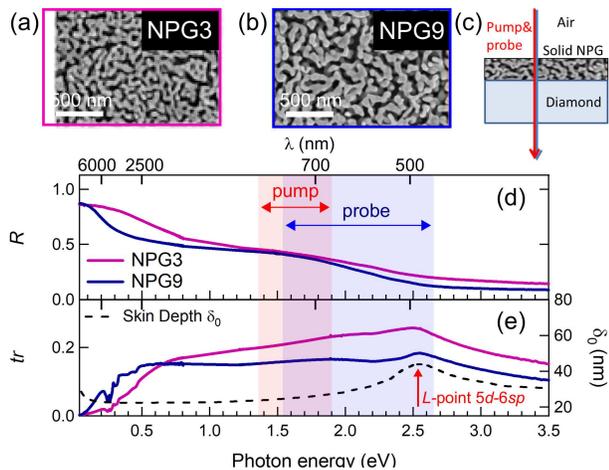}
\caption{(a), (b) Scanning electron micrographs (SEM) of the two NPG samples characterized by different $f$ and $d_{\mathrm{wire}}$. (c) Scheme of the solid thin-film samples with optical beams. (d),(e) Reflectance and transmittance spectra of the NPG films at equilibrium. The transmission dip around \SI{0.3}{eV} in panel (e) is due to multi-phonon absorption in the diamond substrate. In panel (e) the skin depth of gold taken from Ref. \cite{olmon2012optical} is also reported to highlight the dielectric resonance of gold at $2.5$ eV mainly due to 5$d$-6$sp$ interband transition at the $L$-point of the first Brillouin zone.}
\label{fig:Fig1}
\end{figure}

\begin{figure*}[ht]
\centering
\includegraphics[width=0.75\textwidth]{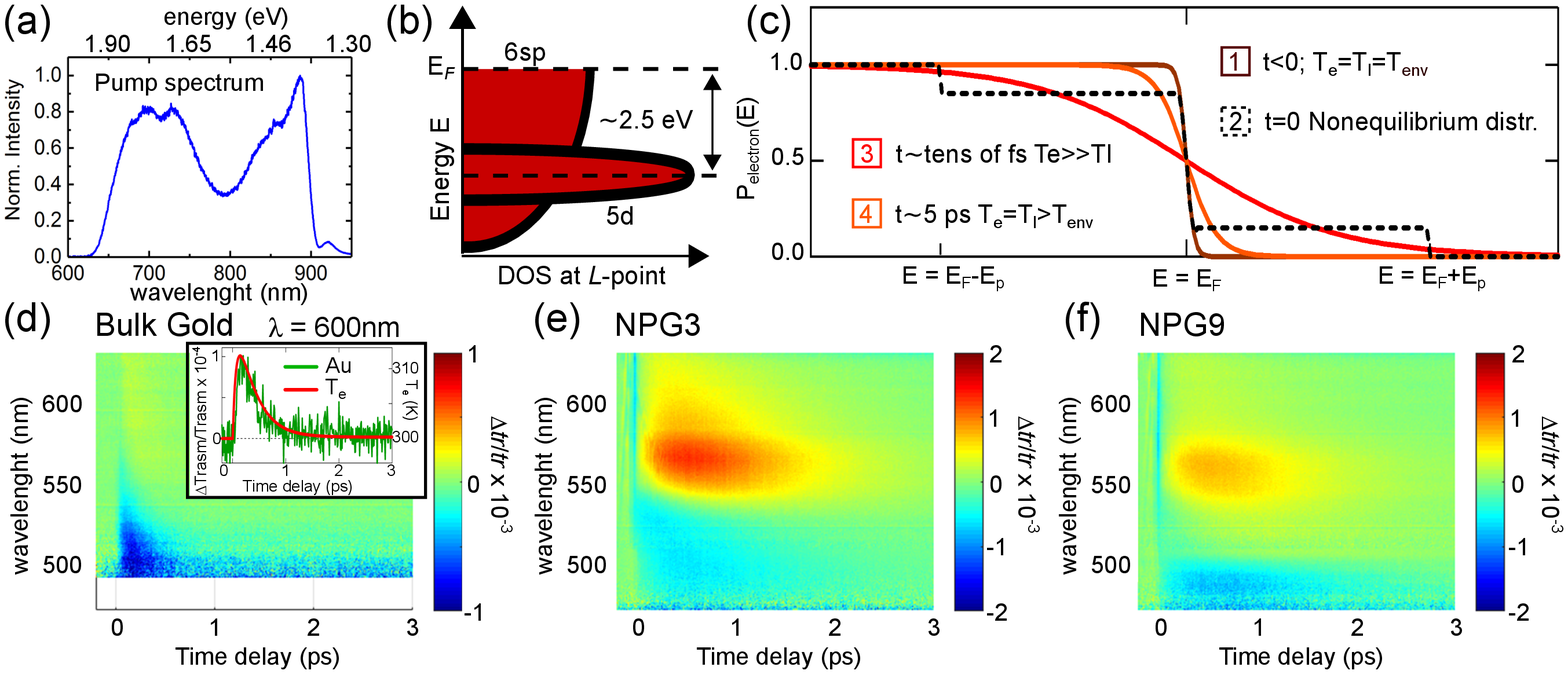}
\caption{(a) Spectrum of the pump pulse used in the experiments (duration is \SI{7}{fs}). (b) Simplified sketch of the density of states (DOS) of gold at the $L$-point employed in this work for interpretation of the pump-probe data. (c) Simplified sketch of the evolution of the Fermi-Dirac distribution following the pump pulse excitation. The shift of the chemical potential with temperature is neglected for clarity. (d-f) $\Delta tr(t)/tr$ maps for a reference bulk gold thin film (thickness \SI{30}{nm}) (d) and for the two NPG samples (e),(f). Inset of panel (d), green curve: cut of the map in (d) at $\lambda=\SI{600}{nm}$; red curve: the $T_{\mathrm{e}}(t)$ obtained from the extended TT model.}
\label{fig:Fig2}
\end{figure*}

NPG \cite{biener2008nanoporous,ding2009nanoporous,lang2011localized,detsi2014localized} represents an interesting system for applications, as it allows liquid and gas samples to fill the 
empty spaces among gold ligaments \cite{lee2012plasmonic,mukherjee2012hot,zhang2017surface,zhang2018plasmonic} where the radiation field is strongly enhanced by cusp-like geometries of the fractal structure (see Fig. \ref{fig:Fig1} (a)-(b)) \cite{lang2011localized,detsi2014localized,garoli2018fractal}. 
Nanoporous materials of different kinds (e.g. glass \cite{costescu2002thermal}, silicon \cite{hopkins2010reduction,wang2011thermal} and polymers \cite{costescu2002thermal}) are also well known for their 
thermal and acoustic insulation properties. The nanoporous structure should then impact on the ultrafast electron temperature dynamics following the absorption of optical energy by plasmons in NPG. 
If compared to bulk gold, the decrease in the thermal conductivity at the interior of the effective material constituted by the nanoporous metal should then lead to higher maximum temperatures and slower local energy relaxation, in a way similar to what observed in gold nanoparticles \cite{AtwaterPRL2017} and clusters \cite{hrelescu2010dna}. In this work, we present an ultrafast pump-probe spectroscopy study and related thermal modeling of plasmon energy relaxation in NPG. Interestingly, relevant fundamental quantities of the TT model such as the speed of sound, the 
Debye temperature and the electron-phonon coupling constant are found to follow a simple power scaling law with the metal filling fraction $f$ in NPG, which quantitatively explains both the longer time scales and the higher electron temperatures observed in our experiments.


NPG samples were prepared by chemical dealloying from an Ag\textsubscript{0.67}Au\textsubscript{0.33} thin film following the procedure reported in Ref. \cite{garoli2018fractal}. The two films studied in this work are characterized by different dealloying times (3 hours for NPG3 and 9 hours for NPG9) and have a similar $f$ (mainly related to the composition of the initial alloy). Different dealloying times lead to different average diameter of the gold ligaments $d_{\mathrm{wire}}$ \cite{garoli2018fractal}. In particular, by numerical analysis of the SEM images of Fig. \ref{fig:Fig1}(a),(b) \cite{garoli2018fractal}, we found $f = 0.39$ and $d_{\mathrm{wire}}\sim \SI{50}{nm}$ for NPG3, $f = 0.37$ and $d_{\mathrm{wire}}\sim \SI{80}{nm}$ for NPG9. In Fig. \ref{fig:Fig1}(c),(d) the optical reflectance $R$ and transmittance $tr$ of the two NPG films in the infrared and visible ranges are reported. A redshift of the plasma edge is observed from \SI{0.5}{eV} in NPG3 to \SI{0.2}{eV} in NPG9 \cite{lang2011localized,detsi2014localized,garoli2018fractal}. The dielectric resonance of gold at \SI{2.5}{eV} is clearly visible in all samples. The broad peak barely seen in the spectra of NPG9 around \SI{1.8}{eV} is due to an effective medium resonance \cite{garoli2018fractal}.

In a simplified model of optical excitations of gold, the lowest-energy interband transition is the 5$d$-6$sp$ transition at the L-point, which leads to the lowest-energy resonance in the dielectric function of gold. The spectral lineshape of this resonance is a Lorentz function centered at \SI{2.5}{eV} \cite{della2012real,AtwaterPRL2017,olmon2012optical}. In this work, in order to focus on the geometrical effect of the nanoporous structure rather than on the details of electromagnetic interactions, we will make use of a corresponding simplified model for ultrafast pump-probe spectroscopy of gold: the infrared pump pulse spectrum, being located at photon energies well below the L-point transitions at \SI{2.5}{eV} (see Fig. 2(a)), mainly excites the intraband transitions within the 6$sp$ band. As a 6$sp$ intraband transition of gold can be seen as a pure free-electron excitation, it can also be interpreted as a plasmon excitation. The plasmon then decays into a 6$sp$ electron-hole pair that subsequently thermalizes in a hot carrier population in the 6$sp$ band, which we model with a simple Fermi-Dirac distribution thermalized at $T_{\mathrm{e}}$. The white-light probe pulse, instead, encompasses a broader spectral range including the dielectric function resonance at \SI{2.5}{eV}, here used as a qualitative probe of $T_{\mathrm{e}}$ as a function of pump-probe delay. Fig. 2(b) is a sketch that summarizes the simplified model for ultrafast pump-probe spectroscopy of gold. However, it has been recently established, both theoretically \cite{AtwaterNatComm2014, AtwaterACSNano2016} and experimentally \cite{AtwaterNatComm2018}, that 5$d$-6$sp$ interband transitions at the $X$-point can actually be excited by pump photons with energy higher than a threshold approximately set at \SI{1.8}{eV}. The effect of $X$-point transitions is to depress plasmon excitation in the 6$sp$ band taking place at pump photon energies higher than \SI{1.8}{eV}, therefore the simplified picture described above and sketched in FIg. 2(b), which implies pure plasmonic excitation in gold for all pump photon energies below the dielectric function resonance at \SI{2.5}{eV}, has to be rigorously rejected \cite{AtwaterNatComm2018}. At odds with the $L$-point transitions, however, the weaker $X$-point transition oscillator does not produce a true resonance in the dielectric function of gold at \SI{1.8}{eV} \cite{olmon2012optical} so our probe pulse will not be sensitive to hot holes in the 5$d$ band at that energy. Also, the pump pulse spectrum in our experiment extends between \SI{1.4}{eV} to \SI{1.9}{eV} as shown in Fig. 2(a), so it overlaps only marginally with the $X$-point transitions at \SI{1.8}{eV}. Therefore, the simplified model of Fig. 2(b) can be fairly employed for the scopes of the present work hence allowing us to describe the electron system, after e-h pair thermalization, with the single parameter $T_{\mathrm{e}}$.

Transient absorption experiments were performed with an ultrafast laser system based on a Yb:KGW regenerative amplifier operating at a repetition time of \SI{20}{\micro s}. A home-built noncollinear optical parametric amplifier delivers excitation pulses with a bandwidth of \SI{0.53}{eV} at a central energy of $E_{\mathrm{p}} \sim \SI{1.65}{eV}$ as reported in Fig. \ref{fig:Fig2}(a) hence excluding the 5$d$-6$sp$ transition (see Fig. \ref{fig:Fig2}(b)). Dielectric chirped mirrors compress the pulses to a duration of \SI{7}{fs}. In Fig. \ref{fig:Fig2}(c) the evolution of the Fermi-Dirac distribution following the excitation of the pump pulse is sketched. At $t=0$ the pulse excites a non-equilibrium distribution whose shape is determined by the pulse energy spectrum in Fig. \ref{fig:Fig2}(a), which can be roughly approximated by a multiple step function (black dashed curve in Fig. \ref{fig:Fig2}(c)) \cite{Govorov,della2012real}. The non-equilibrium e-h pair distribution generated by the pump pulse thermalizes to a Fermi-Dirac distribution at $T_{\mathrm{e}}$ on a timescale of the order of hundreds of fs, mainly through electron-electron interactions. At this stage, $T_{\mathrm{e}}$ is still much higher than $T_{\mathrm{l}}$ (red curve in Fig. \ref{fig:Fig2}(c)). On a longer timescale on the order of ps, the carriers cool down through electron-phonon interactions to a new lattice temperature $T_{\mathrm{l}}=T_{\mathrm{e}}$ (orange curve in Fig. \ref{fig:Fig2}(c)) higher than the environment temperature $T_{\mathrm{env}}\simeq\SI{300}{K}$.

The pump-induced optical transmission change $tr(t)$ is probed by a synchronous white light pulse obtained from supercontinuum generation in a \SI{2}{mm} thick sapphire crystal \cite{grupp2017broadly}. Probe pulses cover a spectral range between 1.55 and \SI{2.64}{eV} including the 5$d$-6$sp$ transition. Spectra of subsequent probe pulses are used to calculate the differential transmission signal $\Delta tr(t)/tr = [tr(t)-tr(t\alt 0)]/tr(t\alt 0)$ with a modulation of the excitation pulses at half the repetition rate. In  Fig. \ref{fig:Fig2}(d)-(f), color plots of $\Delta tr(t)/tr$ as a function of pump-probe time delay $t$ and probe wavelength $\lambda$ are shown for a reference bulk gold thin film and for the NPG3 and NPG9 samples. By comparing the three plots of  Fig. \ref{fig:Fig2}(d)-(f), one immediately sees a strongly increased transmittance around $\lambda=\SI{560}{nm}$ in both NPG samples which is almost absent in the bulk gold film \cite{lang2011localized,detsi2014localized}, accompanied by a decay of $\Delta tr(t)/tr$ slower than that of the gold film at all wavelengths. For probe wavelenghts shorter than $\sim$\SI{550}{nm} the sign of $\Delta tr(t)/tr$ changes to negative because of pump-induced interband absorption \cite{rotenberg2009tunable,rotenberg2010ultrafast,baida2011ultrafast,della2015self,di2018time}. High-energy non-thermalized carriers impact on the transmittance of gold films and nanostructures only for $t \ll \SI{0.5}{ps}$ \cite{della2012real,della2015self}. The transmittance dynamics for probe delays above \SI{0.5}{ps}, instead, can be almost entirely attributed to thermalized carriers and to changes in their $T_{\mathrm{e}}$, displaying a relaxation time scale independent on the probe wavelength \cite{qiu1992short}. In this perspective the strongly increased transmittance observed in NPG (positive areas in Fig. \ref{fig:Fig2}(e),(f)) indicates a much higher value of $T_\mathrm{e}^{\mathrm{Max}}$ if compared to that reached in bulk gold (Fig. \ref{fig:Fig2}(d)). These facts demonstrate that NPG is a very promising candidate for hot-electron plasmonics applications.

\begin{figure*}[ht]
\centering
\includegraphics[width=0.75\textwidth]{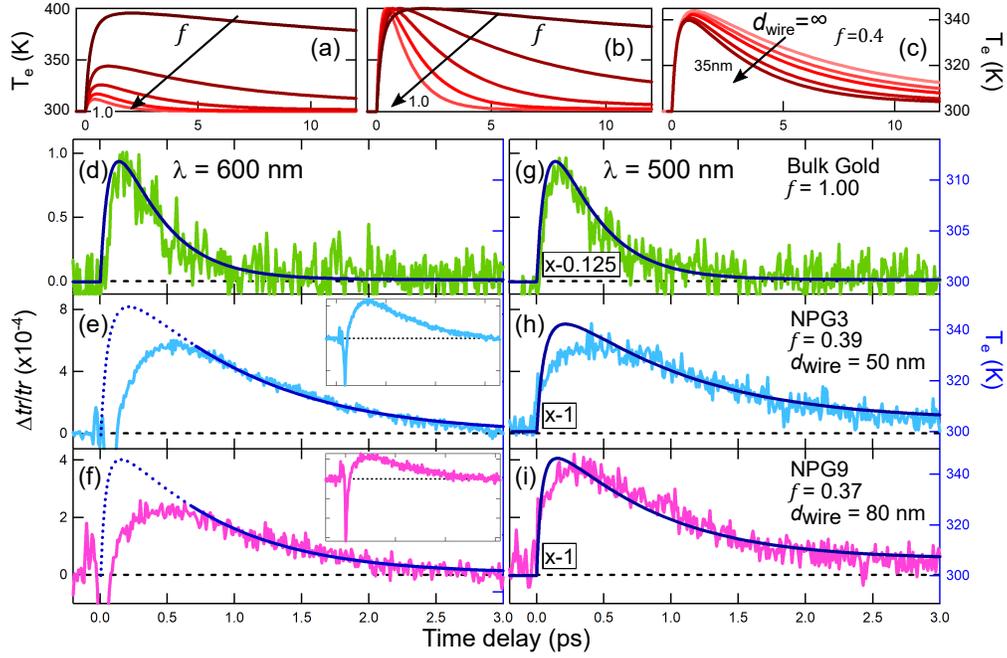}
\caption{(a) Effect of varying $f$ on the $T_\mathrm{e}(t)$ dynamics ($f=1$ corresponds to bulk gold). (b) Same curves as (a) normalized at $T_\mathrm{e}^{Max}$ to highlight the different temperature dynamics. (c) Effect of $d_{\mathrm{wire}}$ on the $T_\mathrm{e}(t)$ decay for $f=0.4$. (d-i) Comparison of the spectra obtained from the $\Delta tr(t)/tr$ color plots of Fig. \ref{fig:Fig2} at $\lambda=\SI{600}{nm}$ (d)-(f) and at $\lambda=\SI{500}{nm}$ (g)-(i) with the electron temperatures obtained from the extended TT model (dark blue curves). (d),(g): bulk gold; (e),(h): NPG3; (f),(i): NPG9. Inset of panels (b),(c) show the full undershoot at short delay  due to the generation of non-thermalized carriers in NPG. Note that $\Delta tr(t)/tr$ in panels (g)-(i) is reported with negative multiplication factors.}
\label{fig:Fig3}
\end{figure*}

Numerical evaluation of $T_\mathrm{e}(t)$ and $T_\mathrm{l}(t)$ dynamics is performed within the two-temperature model, in which energy relaxation to the lattice from the free carriers, heated by e-h pair thermalization via the fast electron-electron interaction, is mediated by the relatively slow electron-phonon interaction \cite{anisimov1974electron}. In an improved version of the TT model \cite{carpene2006ultrafast}, e-h pairs produced by plasmon decay act as the external heat source for both the Fermi-Dirac free carrier distribution and the lattice via electron-electron and electron-phonon scattering processes respectively, resulting in the following coupled equations:
\begin{widetext}
\begin{equation} \begin{split} \label{eq1} C_{\mathrm{e}}\frac{dT_\mathrm{e}}{dt}=-g(T_\mathrm{e}-T_\mathrm{l} )-\frac{e^{-\left (\tau_{\mathrm{e,relax}}^{-1}+\tau_{\mathrm{p,relax}}^{-1}\right )t}}{t^2}\left [t+\tau_{\mathrm{e,relax}}\left (1-e^{t/\tau_{\mathrm{e,relax}}} \right)\right] \cdot P_a \\
C_{\mathrm{l}}\frac{T_\mathrm{l}}{dt}=g(T_\mathrm{e}-T_\mathrm{l} )-\frac{e^{-\left (\tau_{\mathrm{e,relax}}^{-1}+\tau_{\mathrm{p,relax}}^{-1}\right )t}}{t \tau_{\mathrm{p,relax}}} \left [\tau_{\mathrm{e,relax}}\left (1-e^{t/\tau_{\mathrm{e,relax}}} \right)\right] \cdot P_a\end{split}  \end{equation} \end{widetext}
where $C_{\mathrm{e}}$ and $C_{\mathrm{l}}$ are the electronic and lattice heat capacities per unit volume, $g$ is the electron-phonon coupling constant, $\tau_{\mathrm{e,relax}}$ and $\tau_{\mathrm{p,relax}}$ are characteristic times related to the electron-electron and electron-phonon energy relaxation \cite{carpene2006ultrafast}. The pump pulse power in the instantaneous pump-pulse approximation is $P_a=F_a/d$, with $d$ the film thickness, $F_a=(1-R-tr)F$ and $F=\SI{180}{\micro J/cm^2}$ the pump fluence. For bulk gold thin films, the values of the parameters used in the extended TT model are $C_{\mathrm{e}}=\gamma T_\mathrm{e}$, $\gamma=\SI{68}{J m^{-3} K^{-2}}$, $C_{\mathrm{l}}=2.5\cdot \SI{e6}{Jm^{-3} K^{-1}}$, $g=2.2\cdot \SI{e16}{Wm^{-3}K^{-1}}$, $E_{\mathrm{F}}=\SI{7.3}{eV}$, $\tau_{\mathrm{e,relax}}=\SI{136}{fs}$, $\tau_{\mathrm{p,relax}}=\SI{1650}{fs}$, $E_{\mathrm{P}}=\SI{1.65}{eV}$ \cite{della2012real}. In the inset of Fig. \ref{fig:Fig2}(c), the $T_\mathrm{e}$ curve obtained from Eq.\ref{eq1} fits to the $\Delta tr(t)/tr(0)$ data for bulk gold, provided that the delay scale is normalized by the relative change factor $\xi=\mathrm{ln}(\Delta tr^{\mathrm{Max}}/tr)/\mathrm{ln}(\Delta T_\mathrm{e}^{\mathrm{Max}}/T_\mathrm{e}(t\alt0))\simeq 3$.


In order to analyze the ultrafast temperature dynamics of NPG within the extended TT model, we scale all quantities of Eq.\ref{eq1} by $f^{\beta}$, where $\beta$ is the corresponding scaling exponent, as summarized in Table 1. For $C_{\mathrm{e}}$ and $C_{\mathrm{l}}$, the scaling exponent is a trivial $\beta_C=1$ as they scale linearly with the mass density. For the thermal conductance, the problem is considerably more complex due to the NPG network connectivity. Previous works have employed the Asymmetric Bruggeman Theory (ABT) \cite{mclachlan1990electrical} to calculate the electron thermal conductivity in NPG \cite{hopkins2008thermal, makinson1938thermal} and the lattice thermal conductivity of nanoporous glass \cite{costescu2002thermal}. In both cases, the results point toward an experimental value of $\beta_k=3/2$ for thermal conductivities of nanoporous solids. The lattice thermal conductivity is written as $k_l=1/3 C_{\mathrm{l}} v_s l_{ph}$, where $C_{\mathrm{l}}$ is the lattice specific heat, $v_s$ is the speed of sound and $l_{ph}$ is the phonon mean free path. Since $l_{ph}$ and $C_{\mathrm{l}}$ are microscopic quantities that should not depend on the geometry, $v_s$ should scale with the exponent $\beta_v=\beta_k=+3/2$ as well \cite{costescu2002thermal}. There are two quantities in the TT model of Eq.\ref{eq1} that depend on $v_s$. The first quantity is $g$ \cite{kaganov1957sov}:

\begin{equation} \label{eq2} g=\frac{\pi^2 m_e n_e v_s^2}{6T_\mathrm{e}\tau(T_\mathrm{e}, T_\mathrm{l} )} \end{equation}

where $n_e$ is the microscopic electron density, $m_e$ is the electron mass, and $\tau(T_\mathrm{e}, T_\mathrm{l} )$ is the total electron scattering time including electron-electron $\tau_{\mathrm{ee}}$ and electron phonon $\tau_{\mathrm{ep}}$ scatterings. Following Matthiessen's rule and assuming momentum-independent scattering, the effect of electron  scattering at physical boundaries in NPG ligaments can be included in the model by considering an additional scattering time $\tau_{\mathrm{B}}=v_{\mathrm{F}}/d_{\mathrm{wire}}$, where $v_{\mathrm{F}}=1.40\cdot\SI{e6}{m/s}$ is the Fermi velocity in gold \cite{hopkins2008thermal}:
\begin{equation} \label{eq3}\frac{1}{\tau(T_\mathrm{e}, T_\mathrm{l} )} =\frac{1}{\tau_{\mathrm{ee}}} +\frac{1}{\tau_{\mathrm{ep}}}+\frac{1}{\tau_{\mathrm{B}}} =AT_\mathrm{e}^2+BT_\mathrm{l}+\frac{v_{\mathrm{F}}}{d_{\mathrm{wire}}} \end{equation}
In Eq.\ref{eq3} $A$ and $B$ are temperature-independent coefficients that in gold can be taken equal to $A=1.2\cdot \SI{e7}{K^{-2} s^{-1}}$, $B=1.23\cdot \SI{e11}{K^{-1} s^{-1}}$ \cite{wang1994time}. In bulk gold $d_{\mathrm{wire}}\to \infty$ and the contribution of $\tau_{\mathrm{B}}$ is negligible. The case of gold nanoparticles can also be obtained by using $f=1$ and $d_{\mathrm{wire}}$ similar to the value of the nanoparticle diameter \cite{supplmat}. In Eq.\ref{eq2} the only quantity that scales with $f$ is the speed of sound $v_s$, therefore for $g$ we obtain a scaling exponent $\beta_g=2\beta_v=+3$.


The second quantity of the TT model proportional to $v_s$ is the Debye temperature $\Theta_{\mathrm{D}}$:
\begin{equation} \label{eq4} \Theta_{\mathrm{D}}=\frac{\hbar k_{\mathrm{D}} v_s}{k_\mathrm{B}}  \end{equation}
where $k_{\mathrm{D}}=(6\pi N_a)^{1/3}$ ($N_a$ is the atomic density) and $k_\mathrm{B}$ is the Boltzmann constant. $k_\mathrm{B} \Theta_{\mathrm{D}}$ represents the average phonon energy and, as such, enters in the definition of the electron-phonon energy relaxation time as $\tau_{\mathrm{p,relax}}=\tau_{\mathrm{ep}} E_{\mathrm{P}}/k_B \Theta_{\mathrm{D}}$. Therefore, from $\beta_{\Theta}=\beta_v=+3/2$ we obtain $\beta_{\tau}=-\beta_{\Theta}=-3/2$ for $\tau_{\mathrm{p,relax}}$.

\begin{table}[h]
\caption{\label{tab:1}Geometrical scaling of the TT model parameters.}
\centering
\begin{ruledtabular}
\begin{tabular}{lccccc}
Quantity & $C_{\mathrm{e}}, C_{\mathrm{l}}$ & $v_s$ & $g$ & $\Theta_{\mathrm{D}}$ & 
$\tau_{\mathrm{p,relax}}$ \\
\hline
scaling ($f<1$) & $f^1$ & $f^{3/2}$ & $f^{3}$ & $f^{3/2}$ & $f^{-3/2}$ \\
\end{tabular}
\end{ruledtabular}
\end{table}
Using the scaling exponents of Table \ref{tab:1}, we can describe the ultrafast electron dynamics of NPG by solving the extended TT model of Eq.\ref{eq1} as a function of $f$. It is important to notice that the scaled quantities are effective quantities purposely defined for the nanoporous solid, and do not correspond to an actual variation of the microscopic quantities of bulk gold. In Fig. \ref{fig:Fig3}(a)-(c) the model results are reported, highlighting the effect of $f$ and $d_{\mathrm{wire}}$ on $T_\mathrm{e}$. In the model, the temperature dynamics is clearly slowed down for low $f$ and $T_\mathrm{e}^{\mathrm{Max}}$ is considerably increased. Electron scattering at physical boundaries, which is almost absent in bulk gold, becomes relevant only when the electron mean free path in gold $\ell \sim \SI{40}{nm}$ \cite{hopkins2008thermal,gall2016electron} is of the same order of the mean ligament diameter $d_{\mathrm{wire}}$ (as it is in our samples NPG3 and NPG9 with $d_{\mathrm{wire}}$ of $\SI{50}{nm}$ and $\SI{80}{nm}$, respectively).

In Fig. \ref{fig:Fig3}(d)-(i) we compare cuts of the experimental data of Fig. \ref{fig:Fig2}(d)-(f) at fixed $\lambda=\SI{600}{nm}$ and $\lambda=\SI{500}{nm}$ with the prediction of the TT model scaled by $f=0.39$ for NPG3 and $f=0.37$ for NPG9. The relaxation dynamics for $t>\SI{0.5}{ps}$ is fairly reproduced by the TT model in all plots of Fig. \ref{fig:Fig3}. The much higher $\Delta tr(t)/tr$ for NPG if compared to bulk gold at $\lambda=\SI{600}{nm}$ is indicative of the much higher $T_{\mathrm{e}}^{\mathrm{Max}}$ reached in NPG. The TT model accounts only for the dynamics of thermalized electrons and therefore it cannot reproduce the ultrafast variations of $\Delta tr(t)/tr$ at very short $t \geq 0$. Especially at $\lambda=\SI{600}{nm}$, a strong induced absorption signal can be seen for $t<\SI{100}{fs}$ (see insets of panels (e) and (f)) and it can be attributed to the excitation of non-thermalized high-energy carriers \cite{rotenberg2009tunable,rotenberg2010ultrafast,baida2011ultrafast,della2015self,di2018time}. Hot carriers are almost absent in bulk gold for the same excitation conditions as for NPG, as expected due to the high density of field-enhancement hotspots in NPG and to the high surface/volume ratio \cite{Govorov} of the NPG fractal structure \cite{garoli2018fractal}. At $\lambda=\SI{500}{nm}$ the contribution of non-thermalized carriers to $\Delta tr(t)/tr$ is much smaller \cite{della2012real,AtwaterPRL2017} and it does not impact on the fitting of the model to the data as seen in Fig. \ref{fig:Fig3}(h),(i). It has been observed \cite{aruda2013identification} that surface functionalization of gold nanostructures leads to similar slowdown of the temperature dynamics. Further studies of functionalized nanoporous gold for future hot electron chemistry applications will be required to understand the combination of the two different slowdown effects.

In conclusion, the predictions of a geometrical scaling theory of nanoporous gold, concerning the reduced thermal capacitance, the weaker thermal link between electrons and phonons and the longer electron-phonon energy relaxation time if compared to bulk gold, could quantitatively account for the ultrafast temperature dynamics experimentally observed by pump-probe spectroscopy. On the basis of these results, higher electron temperatures and longer plasmon decay times can be engineered in gold nanostructures for future applications of hot-electron plasmonics.


%

\end{document}